\UseRawInputEncoding 

\documentclass[twocolumn,showpacs,floats,floatfix,superscriptaddress,aps,pra]{revtex4-1}

\usepackage{amsfonts}
\usepackage{amssymb}
\usepackage{amsmath}
\usepackage{calc}
\usepackage{graphicx}
\usepackage{bm}

\usepackage[normalem]{ulem}
\usepackage{amsmath,amssymb}
\usepackage{multirow}
\usepackage{xcolor,soul}
\usepackage{srcltx}
\usepackage{hyperref,graphicx}

\def\be{ \begin{equation}}
\def\ee{ \end{equation}}
\def\bea{ \begin{eqnarray}}
\def\eea{ \end{eqnarray}}
\def\bse{ \begin{subequations}}
\def\ese{ \end{subequations}}
\def\bc{ \begin{center}}
\def\ec{ \end{center}}

\newcommand{\stef}[1]{{\color{black} #1}}

\begin{document}

\author{Stefano Longhi$^{*}$} 
\affiliation{Dipartimento di Fisica, Politecnico di Milano, Piazza L. da Vinci 32, I-20133 Milano, Italy}
\affiliation{IFISC (UIB-CSIC), Instituto de Fisica Interdisciplinar y Sistemas Complejos, E-07122 Palma de Mallorca, Spain}
\email{stefano.longhi@polimi.it}

\title{Non-Hermitian Maryland Model}
  \normalsize


%
\bigskip
\begin{abstract}
\noindent  
Non-Hermitian systems with aperiodic order display phase transitions that are beyond the paradigm of Hermitian physics.
This motivates the search for exactly solvable models, where localization/delocalization phase transitions, mobility edges in complex plane and their topological nature can be unraveled. 
Here we present an exactly solvable model of quasi crystal, which is a non-pertrurbative non-Hermitian extension of a famous integrable model of quantum chaos proposed by Grempel {\it at al.} [Phys. Rev. Lett. {\bf 49}, 833 (1982)] and dubbed the  Maryland model.  Contrary to the Hermitian Maryland model, its non-Hermitian extension shows a richer scenario, with a localization-delocalization phase transition via topological mobility edges in complex energy plane.
 \end{abstract}

\maketitle

\section{Introduction}
Topological phases and localization-delocalization phase transitions in non-Hermitian synthetic matter with periodic or aperiodic order have recently sparked tremendous interest in different areas of physics, from condensed matter physics to cold atoms and classical systems \cite{r1,r2,r3,r4,r5,r5b}. Non-Hermitian crystalline systems  display a variety of exotic effects, such as non-trivial topology even in one-band systems, the non-Hermitian skin effect and the breakdown of the bulk-boundary correspondence based on Bloch band topological invariants \cite{r6,r7,r8,r9,r10,r11,r12,r13,r14,r15,r16,r17,r18,r19,r20,r21,r22,r23,r24,r25,r26,r27,r28,r29,r30,r31,r32,r33,r33b,r34,r35,r36,r36b,r37,r38,r39,r40,r41,r42,r43,r44,r45,r45b,r46,E1,E2,E3,E4,E5,E6,r47,r47a,r47b,r47c}. In systems with aperiodic order (quasi crystals), non-Hermiticity induces phase transitions that are beyond the paradigm of Hermitian quasi crystals \cite{r48,r49,r50,r51,r52,r53,r54,r55,r56,r57,r58,r59,r60,r60a,r60b}.
 Recent works focused on several non-Hermitian extensions of the famous Aubry-Andr\'{e}-Harper model \cite{A1,A2,A3}, showing how localization-delocalization phase transitions and mobility edges in complex energy plane, separating extended and  localized states, can be characterized by point-gap topological numbers \cite{r1,r50,r52,r53,r54,r55,r57,r58,r59,r60}.
While non-Hermitian models of crystalline systems are exactly integrable using a suitable non-Hermitian extension of Bloch band theory \cite{r9,r10,r19,r35}, in a quasi crystal the incommensurability of the underlying potential makes the problem not integrable in most cases, even in the Hermitian limit. An exception is provided by a famous model of localization in quantum chaos, introduced by Grempel {\it et al.} \cite{r61} in connection to Anderson localization in periodically-kicked quantum systems \cite{r62,r63,r63b,r64,r64b,r64c,r64d,r64e,r65,r66,r67,r67b} and dubbed the "Maryland" model by Barry Simon \cite{r64b}.  Both the Aubry-Andr\'{e}-Harper and Maryland models can be associated to a two-dimensional lattice with integer quantum Hall topology \cite{r69}. However, only the latter model is integrable.
In fact, the Aubry-Andr\'{e}-Harper model displays self-duality, useful to determine the phase transition point and Lyapunov exponent, however it is not possible to 
solve the model in terms of its spectrum. Conversely, the Maryland model is integrable after mapping the spectral problem of the incommensurate potential into an integrable dynamical (Floquet) problem.\\ 
In this work we introduce a  non-perturbative non-Hermitian (NH) extension of the Maryland model and show that this model is integrable, thus providing a rather unique example of exactly-solvable NH quasi crystal. Unlike its Hermitian limit, the NH Maryland model shows a localization-delocalization phase transition with topological mobility edges in complex energy plane.\\
\par

\section{Non-Hermitian Maryland model and its energy spectrum}
 We consider the non-perturbative NH extension of the Maryland model \cite{r61,r64b} on a one-dimensional tight-binding lattice with nearest-neighbor hopping  defined by the eigenvalue equation
\begin{equation}
H(\theta, \epsilon) \psi_n \equiv \psi_{n+1}+\psi_{n-1}+V \tan ( \pi \alpha n + \theta + i \epsilon )\psi_n = E \psi_n ,
\label{eq1}
\end{equation}
for the wave function amplitude $\psi_n$ at the $n$th lattice site, where the hopping amplitude is set to unity, \stef{ $\alpha$ is irrational with typical Diophantine properties and measure of irrationality $L(\alpha)=0$ \cite{r64b}}, $V$ and $\theta$ are the amplitude and phase of the incommensurate potential, 
$i \epsilon$ is an additional complex phase that makes the model NH. \stef{A typical irrational with measure $L(\alpha)=0$ is $\alpha=(\sqrt{5}-1)/2$ (inverse of the golden mean)}. In the following we will assume $\epsilon, V \geq 0$ for the sake of definiteness. For a non-vanishing NH parameter $\epsilon$, the complex potential $V_n=V \tan ( \pi \alpha n + \theta + i \epsilon )$ is bounded, showing specially-tailored profiles for its real and imaginary parts. Such a  NH Hamiltonian $H(\epsilon, \theta)$ with tailored potential can be implemented, for example, in synthetic photonic lattices \cite{E5,E6,r55,ruffa}, where the combined use of amplitude and phase modulators in fiber loops enable to synthesize rather arbitrary complex potential profiles.\par
 A rather unique property of $H(\epsilon, \theta)$ in the Hermitian limit $\epsilon=0$ is its integrability \cite{r61,r64b}. Since for $\epsilon=0$ the potential $V_n$ is unbounded, there are not extended states, the energy spectrum $E$ is pure point and dense in the interval $(-\infty,\infty)$, and all wave functions are exponentially localized with an energy-dependent Lyapunov exponent $L(E)$ (inverse of the localization length) given by \cite{r61}
\begin{equation}
 L(E)=  {\rm arcosh} \left\{ \frac{\sqrt{(2+E)^2+V^2}+\sqrt{(2-E)^2+V^2}}{4} \right\}
\end{equation}
which is strictly positive for any energy $E$.\\
When going to the NH case $\epsilon>0$, a few natural questions arise: Is the NH Maryland model integrable?  How does the energy spectrum $E$ is modified by the complex potential? Does the NH Maryland model show topological features rooted in the complex nature of its energy spectrum?\\
Before answering to all such questions, let us first compute the Lyapunov exponent $L(E)$, as a function of the complex energy $E=E_R+iE_I$, for $\epsilon>0$. The calculation of $L(E)$ can be performed analytically using Avila's global theory for quasi-periodic operators \cite{r70,r71}. The result is that $L(E)$ is again given by Eq.(2), but with the replacement $E \rightarrow E_R$ and $V \rightarrow V-E_I$, where $E_R$ and $E_I$ are the real and imaginary parts of $E$ (technical details are given in Appendix A). Interestingly, as $E$ varies on the segment $\Gamma$ of the complex energy plane  defined by $E_I=V$ and  $-2 \leq E_R \leq 2$, one has $L(E)=0$, indicating that in the NH Maryland model extended states, and possibly coexistence of localized and extended wave functions (i.e. mobility edges in complex plane), can be found, unlike its Hermitian limit.
 The exact form of energy spectrum and eigenfunctions can be derived from the integrability of the NH Maryland model, which is presented below. Here we  summarize the main results, which are illustrated in Fig.1. The energy spectrum is independent of the phase $\theta$. As $\epsilon$ increases above zero, there is a transition from a  localized phase with all wave functions exponentially localized for $\epsilon<\epsilon_1$ , to a mobility edge phase with coexistence of localized and delocalized wave functions for $\epsilon_1 < \epsilon < \epsilon_2$, to a delocalized phase with all wave functions delocalized for $\epsilon> \epsilon_2$. Remarkably, the energy spectrum becomes independent of both $\theta$ {\em and} $\epsilon$ for $\epsilon> \epsilon_2$. The critical values $\epsilon_1$ and $\epsilon_2$ are given by 
\begin{eqnarray}
\epsilon_1 & = & \frac{1}{2} {\rm arsinh}(V)  \\
\epsilon_2 & = & \frac{1}{2} {\rm arcosh} \sqrt{ \frac{2+V^2}{2} + \sqrt{ \left( \frac{2+V^2}{2} \right)^2 -1}}.
\end{eqnarray}
In the {\em  localized phase} $\epsilon<\epsilon_1$, the energy spectrum describes a closed loop $\mathcal{C}$ without self-intersections in the upper half of complex $E$ plane [Fig.1(a)], given by
\begin{equation}
E(\omega)=\sqrt{4 \cos^2( \omega + i \epsilon)+V^2 {\rm cotan}^2 (\omega+ i \epsilon)}
\end{equation}
with $ 0 \leq \omega \leq \pi$ and ${\rm Im}(E) \geq 0$. Remarkably, the Hermitian Maryland model is a non-perturbative result since in the $\epsilon \rightarrow 0$ limit the loop $\mathcal{C}$ tends to a circumference of large radius $\sim V/(2 \epsilon)$, centered at $E=iV/(2 \epsilon)$, rather than to the real energy axis. This is due to the divergence of the tan potential in the Hermitian limit, so that even for a small $\epsilon$ the potential change is non-perturbative at some sites in the lattice.\\
In the {\em mobility edge phase} $\epsilon_1<\epsilon < \epsilon_2$, the energy spectrum associated to exponentially-localized eigenstates describes two closed and non-crossing loops $\mathcal{C}_{1,2}$ encircling the two points $E_{\pm}=\pm 2+iV$ in complex energy plane. The curves $\mathcal{C}_{1,2}$ are the two outer sub loops defined by Eq.(5) when $\epsilon_1< \epsilon < \epsilon_2$. The energy spectrum of extended states
is the portion of the $\Gamma$ segment connecting the two loops $\mathcal{C}_1$ and $\mathcal{C}_2$ [Fig.1(b)]. As $ \epsilon$ approaches $\epsilon_2$ from below, the loops $\mathcal{C}_{1,2}$ shrink toward the two points $E_{\pm}$.\\
In the {\em delocalized phase} $\epsilon> \epsilon_2$ the energy spectrum is the entire $\Gamma$ segment in the upper complex energy plane, and remarkably it does not depend on $\epsilon$ anymore.\\

\begin{figure}[t]
   \centering
    \includegraphics[width=0.45\textwidth]{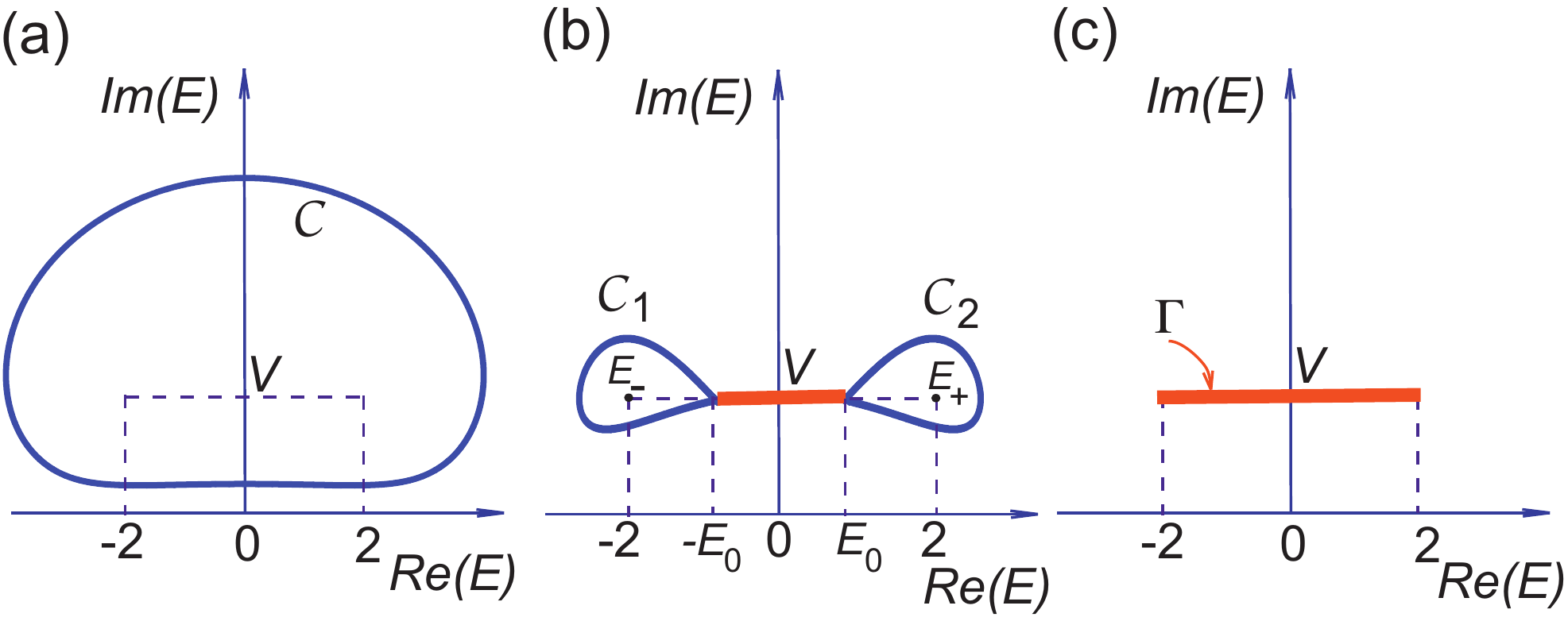}
    \caption{Schematic of the energy spectrum $E$ of the NH Maryland model in the three different phases: (a) localized phase ($\epsilon<\epsilon_1$), (b) mobility edge phase ($\epsilon_1 < \epsilon < \epsilon_2$), and 
    (c) extended phase ($\epsilon> \epsilon_2$). The closed loops $\mathcal{C}$ and $\mathcal{C}_{1,2}$ in (a) and (b) correspond to exponentially-localized eigenstates, wheres the horizontal straight segments in (b) and (c) correspond to extended states. In (b) the real part of energy spectrum of extended states is bounded between $(-E_0,E_0)$, where $E_0=2 \cos \omega_0$ and $\cos \omega_0$ is given by Eq.(14).}
     \label{fig1}
\end{figure}

\section{Integrability of the non-Hermitian Maryland model} To show the integrability of the NH Maryland model, as in Refs.\cite{r61,r63b} we map the static problem (1) into an equivalent dynamical (Floquet) problem, which is exactly solvable. Here we outline the procedure and present the main results, while some technical proofs are detailed in the Appendix B. Let us set 
\begin{equation}
\psi(x)=\sum_{n=-\infty}^{\infty} \psi_n \exp(inx) \;, \;\; u(x)= \psi(x) [1-iW(x)]
\end{equation}
where $x$ is real, and
\begin{equation}
W(x)=-\frac{E}{V}+\frac{2}{V} \cos(x).
\end{equation}
 Physically, $\psi(x)$ --and associated function $u(x)$-- provides a spectral representation of the eigenfunction $\psi_n$ of $H(\theta, \epsilon)$ in Fourier (Bloch) domain. Note that if $E$ does not belong to $\Gamma$ and is an eigenvalue corresponding to a localized state in physical space ($ \sum_n |\psi_n|^2 < \infty$), $1-iW(x) \neq 0$ and $u(x)$ is a continuous function of $x$ with $u(x+2 \pi)=u(x)$, that is the state is extended and periodic in Fourier space. On the other hand, for an extended state $\psi_n$ in physical space the energy $E$ belongs to the segment $\Gamma$ and $u(x)$ has to be regarded as a generalized function, given by a superposition of  delta Dirac combs and corresponding to localization in Fourier space \cite{r63b}. In terms of the new function $u(x)$, the discrete equation (1) is mapped into the recurrence equation
 (see Appendix B)
\begin{equation}
u(x-2 \pi \alpha)=g(x) u(x)
\end{equation}
where we have set
\begin{equation}
g(x)= \exp(- 2 \epsilon+2 i \theta) \frac{1+iW(x)}{1-iW(x)}.
\end{equation}
{\it A. Localized states.} Let us first assume that $E$ belongs to the point spectrum of $H(\theta,\epsilon)$, i.e. the wave function $\psi_n$ is localized in physical space. Then $u(x)$ is a continuous and periodic function of $x$ with $2 \pi$ period. Since $\alpha$ is irrational, Eq.(8)  necessarily implies the solvability condition (see Appendix B)
\begin{equation}
\int_0^{2 \pi} dx \ln | g(x) |=0.
\end{equation}
which provides a double constraint on the energy $E$: (i) $E$ should belong to the curve defined by Eq.(5), and (ii) the condition ${\rm Im}(\varphi_+)<{\rm Im}(\varphi_-)<0$ should be satisfied, where $\varphi_{\pm} \equiv {\rm acos} [ (E \pm iV)/2]$. For $\epsilon<\epsilon_1$ [$\epsilon_1$ is defined by Eq.(3)], it turns out that the curve (5) is a closed loop without self-intersections, and the  condition (ii) is always satisfied. Hence for $\epsilon< \epsilon_1$ the entire curve (5), indicated by $\mathcal{C}$ in Fig.1(a), describes the location of energies $E$ in complex plane corresponding to exponentially-localized wave functions. For $\epsilon_1< \epsilon < \epsilon_2$, the curve defined by Eq.(5) is a loop with two self-intersections, which divide the curve into three non-intersecting sub-loops. The condition (ii) is met on the two outer sub loops, denoted by $\mathcal{C}_1$ and $\mathcal{C}_2$ in Fig.1(b). Thus for $\epsilon_1< \epsilon < \epsilon_2$ the energies $E$ corresponding to the localized wave functions lie on the two loops $\mathcal{C}_{1,2}$. Finally, for $\epsilon>\epsilon_2$ the curve defined by Eq.(5) is an open loop in the upper half complex energy plane without self-intersections, however the condition (ii) is not met. Therefore for $\epsilon> \epsilon_2$ the point spectrum of $H(\theta,\epsilon)$ is empty and there are not localized wave functions.\\
{\it B. Extended states.} An extended wave function $\psi_n$ necessarily should correspond to an energy $E$ belonging to $\Gamma$, where the Lyapunov exponent $L(E)$ vanishes. Hence we set $E=i V + 2 \cos \omega$, with $\omega$ real in the range $(0,2 \pi)$. Note that $1-iW(x)$ vanishes at $x=\omega$, and Eq.(8) should be understood as 
\begin{equation}
u(x-2 \pi \alpha) [1-iW(x)]=u(x)[1+i W(x)] \exp( -2 \epsilon+2 i \theta).
\end{equation}
Clearly, for irrational $\alpha$ Eq.(11) cannot be satisfied for any regular function $u(x)$ (unless $u(x) \equiv 0$). The solution to Eq.(11) can be searched as a series of $\delta$-Dirac functions with incommensurate periods $ 2 \pi$ and $2 \pi \alpha$ \cite{r63b}, i.e. 
\begin{equation}
u(x)=\sum_{n,l} U_l \delta(x- \omega-2 \pi \alpha l - 2 \pi n)
\end{equation}
with amplitudes $U_l$ which should vanish as $l \rightarrow \pm \infty$.
 Equation (11) is satisfied for the amplitudes $U_l$  (see Appendix B)
\begin{equation}
U_l= \left\{
\begin{array}{ll}
0 & l  \geq 0 \\
1 & l=-1 \\
\prod_{n=1}^{-l-1} \exp(-2 \epsilon+2 i \theta) \frac{1+iG_{-n}}{1-iG_{-n}}  & l<-1
\end{array}
\right.
\end{equation}
where we have set $G_l \equiv W(x=\omega+2 \pi \alpha l)$. The condition $|U_l| \rightarrow 0$ as $l \rightarrow - \infty$ requires that $|{\rm Im}(\varphi)|<2 \epsilon$, where the complex angle $\varphi$ is defined by $\cos \varphi \equiv (E+iV)/2=iV+ \cos \omega$. One should distinguish three cases. (i) For $\epsilon< \epsilon_1$, the condition $|{\rm Im}(\varphi)|<2 \epsilon$ is not satisfied for any $\omega$ in the range $(0, 2 \pi)$, i.e. there are not extended states. (ii) For $\epsilon_1 <\epsilon< \epsilon_2$, the condition $|{\rm Im}(\varphi)|<2 \epsilon$ is satisfied for $|\omega|< \omega_0$, where
\begin{equation}
\cos \omega_0 =  \cosh (2 \epsilon) \sqrt{1-\frac{V^2}{ \sinh^2 (2 \epsilon)}}.
\end{equation}
This means that, for $ \epsilon_1 < \epsilon < \epsilon_2$, the energies $E=iV+2 \cos \omega$ in the interval $|\omega|< \omega_0$, describing a portion of the segment $\Gamma$, correspond to extended states [Fig.1(b)]. (iii) For $\epsilon> \epsilon_2$, the condition $|{\rm Im}(\varphi)|<2 \epsilon$ is satisfied for any $\omega$, i.e. all energies $E$ on $\Gamma$ correspond to extended states.\\
\par
\begin{figure*}[htbp]
  \includegraphics[width=176mm]{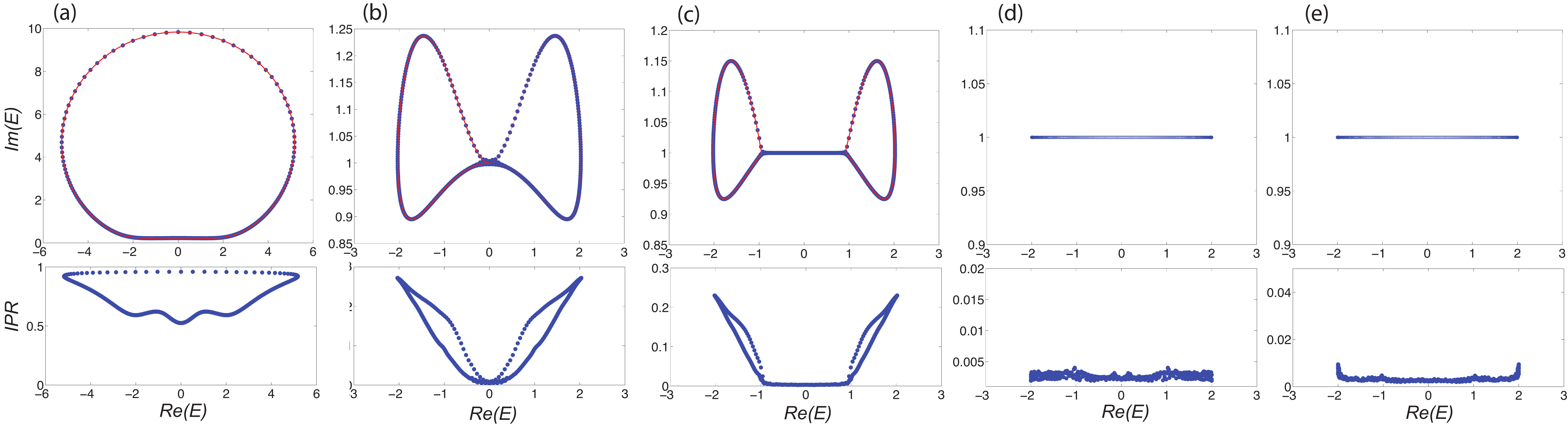}\\
  \caption{(color online) Numerically-computed energy spectrum (upper panels) and IPR of corresponding eigenstates $\psi_n$ (lower panels) of the NH Maryland Hamiltonian $H(\theta,\epsilon)$  in a lattice comprising $L=610$ sites with periodic boundary conditions for $\alpha= (\sqrt{5}-1)/2$, $V=1$, $\theta=0$ and (a) $\epsilon=0.1$, (b) $ \epsilon= \epsilon_1 \simeq 0.4407$, (c) $\epsilon=0.46$, (d) $\epsilon=\epsilon_2 \simeq 0.5306$, and (e) $\epsilon=0.6$. The solid curves in (a), (b) and (c) show the loops $\mathcal{C}$, $\mathcal{C}_1$ and $\mathcal{C}_2$ defined by Eq.(5).}
\end{figure*}
\section{Winding numbers}  The energy spectrum of extended states is a straight segment in complex energy plane, whereas the energy spectrum of localized states describes one or two closed loops, as shown in Fig.1. Therefore, the energy spectrum of localized states can be associated to a winding number $w(E_B)$, for a given base energy $E_B$, given by \cite{r50,r54,r55,r57,r58}
\begin{equation}
w(E_B)=  \lim_{L \rightarrow \infty} \frac{1}{2 \pi i } \int_0^{\pi} d \theta  \frac{\partial}{\partial \theta}  \log  \left\{ \det \left( H \left( \frac{\theta}{L}, \epsilon \right)-E_B \right) \right\}.
\end{equation}
In the calculation of $w(E_B)$, a rational approximation $\alpha \simeq p/L$ of the irrational $\alpha$ is assumed, and $H(\theta, \epsilon)$ is the Hamiltonian of a lattice of size $L$ under periodic (ring) boundary conditions \cite{r50,r58,nota}. When the base energy $E_B$ is internal (external) to either one of the loops $\mathcal{C}$, $\mathcal{C}_1$ or $\mathcal{C}_{2}$ of Fig.1, one has $w(E_B)=1$ ($w(E_B)=0$, respectively). This is because in the Fourier (Bloch) space the localized eigenstates of $H(\theta, \epsilon)$ become extended states and the  phase $\theta$ corresponds to the application of a magnetic flux in the ring, which rotates the position of eigenvalues of the extended states along the closed loops \cite{r1}.  \stef{Such geometric spectral rotations as $\theta$ is varied are shown in Appendix C}. On the other hand, the extended states of $H(\theta, \epsilon)$ become localized states in Fourier space, and their eigenenergies are thus independent of the applied magnetic flux $\theta$ and do not contribute to $w(E_B)$. Such a result can be used to provide a topological characterization of the mobility edge phase \cite{r54,r55}. In fact, if we introduce the two winding numbers $w_1=w(E_B= iV)$ and $w_2=w(E_B=iV \pm 2)$, \stef{the system is in the mobility edge phase when $w_1 \neq w_2$,  whereas it is in the localized or extended phases when $w_1=w_2$.}\\
\par
 \section{Numerical results}  To support the analytical results, we performed numerical calculations of energy spectrum and corresponding eigenstates of the NH Maryland model. We diagonalize the Hamiltonian $H(\theta, \epsilon)$ on a finite lattice with a large number $L$ of sites under periodic boundary conditions. To this aim, the irrational $\alpha=(\sqrt{5}-1)/2$ is approximated by the sequence  $\alpha_n= p_n /p_{n+1}$, where $p_n=0,1,1,2,3,5,8,13,21,34,55, 89, 144, ...$ are the Fibonacci numbers, and take $L=p_n$ for large enough $n$. The inverse participation ratio (IPR) of various eigenstates $\psi_n$, defined by 
\begin{equation}
\text{IPR} =\frac{\sum_{n=1}^{L} \left|\psi_{n}\right|^{4}} {\left( \sum_{n=1}^{L} | \psi_n|^2 \right)^2}
\end{equation}
 is used to distinguish the localized and extended states. The IPR of an extended state scales as $L^{-1}$, hence vanishing in the thermodynamic limit $L \rightarrow \infty$, while it remains finite for a localized state.
In Fig.2, we show the numerically-computed energy spectrum and IPR of the Hamiltonian $H$ for a few increasing values of $\epsilon$, while Fig.3 depicts the behavior of the winding numbers $w_1$ and $w_2$ versus $\epsilon$. The numerical results are in excellent agreement with the analytical results and clearly illustrate the appearance of three different phases, the existence of mobility edges in complex energy plane and their  topological characterization based on the winding numbers associated to the complex spectrum of localized states.\\ 
 \begin{figure}[htbp]
  \includegraphics[width=83mm]{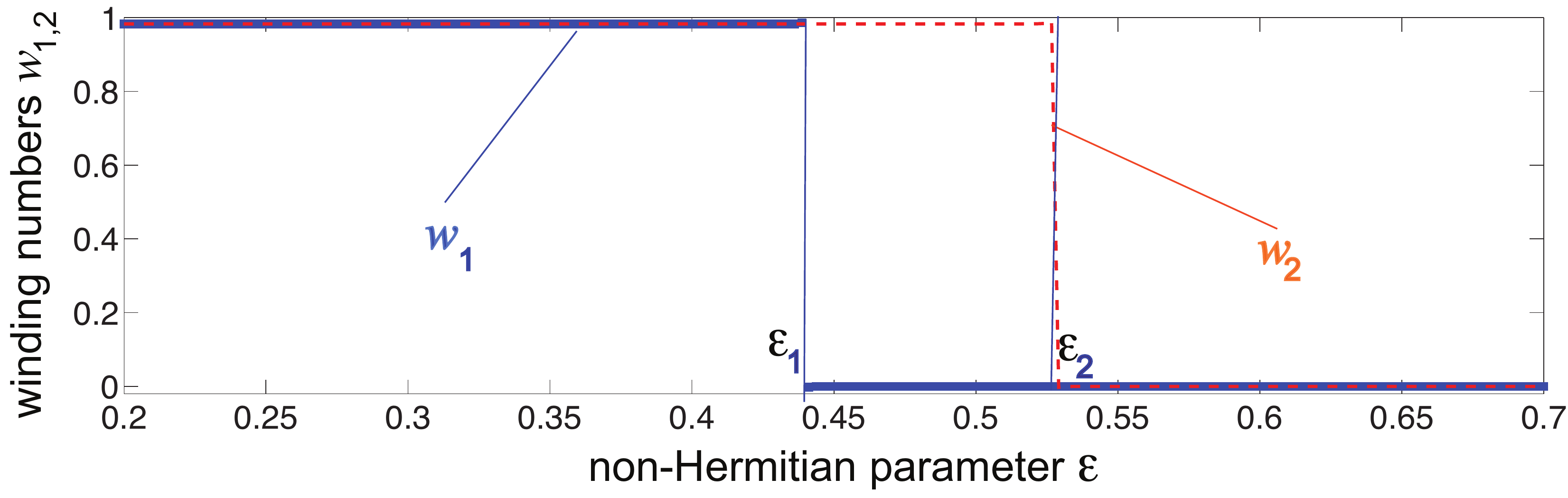}\\
  \caption{(color online) Behavior of the winding numbers $w_1$ and $w_2$ (corresponding to the base energies $E_B=iV$ and $E_B=iV+2$, respectively) versus $\epsilon$ for $V=1$ and $\alpha=(\sqrt{5}-1)/2$. A ring comprising $L=377$ sites has been assumed to compute $w_1$ and $w_2$. The thin vertical lines depict the phase transition points $\epsilon_1$ and $\epsilon_2$, given by Eqs.(3) and (4).}
\end{figure}
\par

\section{Conclusions} In summary, we introduced a non-perturbative and non-Hermitian extension of a famous model of quantum chaos, the Maryland model, and proved its  integrability.
Contrary to the Hermitian Maryland model, its non-Hermitian extension shows a richer scenario, with a localization-delocalization phase transition via topological mobility edges in complex energy plane for irrational $\alpha$ \stef{with zero measure $L(\alpha)=0$}. Our results provide a rather unique example of integrable non-Hermitian system with aperiodic order,  and could have interesting extensions, such as in the study of phase transitions and topological properties of integrable models of two-dimensional NH quasicrystals, and potential relevance in the study of quantum dynamics and quantum chaos in non-Hermitian Floquet systems \cite{C0,C1,C2,C3,C4}. \stef{Finally, it would be of great interest to consider the NH extension of the Maryland model for irrationals with non-zero measure $L(\alpha)>0$ \cite{r64}, such as Liouvillian numbers, where arithmetic phase transitions \cite{r64,r71} could compete with non-Hermitian-driven topological phase transitions.}\\
\\
\acknowledgments
The author acknowledges the Spanish State Research Agency, through the Severo Ochoa
and Maria de Maeztu Program for Centers and Units of Excellence in R\&D (Grant No. MDM-2017-0711).

\appendix
\section{Lyapunov exponent analysis}
The spectral problem
\begin{equation}
\psi_{n+1}+\psi_{n-1}+V_n \psi_n=E \psi_n, \label{S1}
\end{equation}
with $V_n=V \tan (\pi \alpha n+ \theta+i \epsilon)$, can be written in  matrix for as
\begin{eqnarray}
\left(
\begin{array}{l}
\psi_{n+1} \\
\psi_n
\end{array}
\right)=
\left(
\begin{array}{ll}
 E-V_n & -1\\
 1 & 0
\end{array}
\right) 
\left(
\begin{array}{l}
\psi_{n} \\
\psi_{n-1}
\end{array}
\right)  \nonumber \\
\equiv B_n (\theta, \epsilon) \left(
\begin{array}{l}
\psi_{n} \\
\psi_{n-1}
\end{array}
\right).
\end{eqnarray}
For a given complex energy $E$, either belonging or not to the spectrum of $H(\theta, \epsilon)$, the Lyapunov exponent $L(E)$  is given by \cite{r70,r71}
\begin{equation}
L(E)=  \lim_{n \rightarrow \infty} \frac{1}{2 \pi n} \int_0^{2 \pi} d \theta \log || T_n(\theta, \epsilon) ||
\end{equation}
where $\| T_n (\theta) \|$ is the norm of the $2 \times 2$ transfer matrix $T_n(\theta, \epsilon)=B_{n} (\theta, \epsilon) B_{n-1} (\theta, \epsilon)... B_1 (\theta, \epsilon)$.
\stef{Like in the original work by Grempel, Fishman and Prange  \cite{r61,r63b}, we consider here an irrational $\alpha$ with typical Diophantine properties, i.e. we assume that the measure of the irrationality $L(\alpha)$, defined as \cite{r64b} 
\[
L(\alpha)=- \limsup_{n \rightarrow \infty} \frac{1}{n} \log | \sin( \pi \alpha n)|
\]
vanishes, thus excluding from the analysis possible arithmetic phase transitions, i.e. phase transitions that would be observed for irrationals with $L(\alpha)>0$ \cite{r71}.
}

To calculate $L(E)$, we apply Avila's global theory \cite{r70}, extended to the case of a singular potential \cite{r71}, by first calculating $L(E)$ in the $ \epsilon \rightarrow \infty$ limit, and then using the quantization theorem of the acceleration and the properties of convexity, continuity and symmetry of $L(E)$  with respect to $\epsilon$. As shown in 
\cite{r71}, a remarkable property of the Maryland model is that $L(E)$ is independent of $\epsilon$, and thus it can be readily computed in the large $\epsilon$ limit. The main difference here, as compared to \cite{r71}, is that we allow $E$ to be a complex number. In the $\epsilon \rightarrow \infty$ limit, the matrix $B_n( \theta, \epsilon \rightarrow \infty)$, which we indicate by $B^{\infty}$,  turns out to be independent of $n$ and $\theta$, and reads
\begin{equation}
B^{\infty}= \left(
\begin{array}{ll}
E-iV & -1 \\
1 & 0
\end{array}
\right).
\end{equation}
Since $T_n( \theta, \epsilon \rightarrow \infty)=\left( B^{\infty} \right)^n$, $L(E)$ is simply given by $L(E)=\log |\lambda|$, where $\lambda$ is the most unstable eigenvalue of $B^{\infty}$. After introduction of the complex angle $\varphi$ such that 
\begin{equation}
\cos \varphi \equiv (E-iV)/2, \label{S5}
\end{equation}
 the eigenvalues $\lambda_{\pm}$ of the matrix $B^{\infty}$ read $\lambda_{\pm}=\exp( \pm i \varphi)$, and thus $L(E)=| {\rm Im} (\varphi) |$. After setting $E=E_R+iE_I$ and using Eq.(\ref{S5}), after some straightforward algebra one can compute ${\rm Im} (\varphi)$, yielding 
 \begin{widetext}
 \begin{equation}
 L(E)=  {\rm arcosh} \left\{ \frac{ \sqrt{(2+E_R)^2+(V-E_I)^2}+\sqrt{(2-E_R)^2+(V-E_I)^2}}{4} \right\}.
\end{equation}
\end{widetext}
Note that $L(E)$ vanishes whenever $E=iV+2 \cos \omega$  with $\omega$ real, i.e. on the straight segment $\Gamma$ in the complex energy plane shown in Fig.1(c).

\section{Integrability of the NH Maryland Model}


\subsection{The dynamical model}
The integrability of the NH Maryland model can be demonstrated by mapping the spectral problem (\ref{S1}) into an equivalent dynamical (Floquet) problem \cite{r61,r63b}. To this aim, let us set $W_0=-E/V$, $W_1=W_{-1}=1/V$ and $W_l=0$ for $ l \neq 0, \pm1$. Then the eigenvalue equation $H(\theta, \epsilon) \psi_n=E \psi_n$ can be cast in the form
 \begin{equation}
 \sum_l W_l \psi_{n-l}+v_n \psi_n=0 \label{S7}
 \end{equation}
where we have set
\begin{equation}
v_n= \tan ( \pi \alpha n + \theta + i \epsilon). \label{S8}
\end{equation}
Form Eqs.(\ref{S7}) and (\ref{S8}) one has
\begin{equation}
\frac{\psi_n+i \sum_l W_l \psi_{n-l}}{\psi_n-i \sum_l W_l \psi_{n-l}}=\frac{1-iv_n}{1+iv_n}=\exp[-2i ( \pi n \alpha+ \theta + i \epsilon)]. \label{S9}
\end{equation}
Let us formally introduce the series
\begin{equation}
\psi(x)= \sum_n \psi_n \exp(in x) \label{S10}
\end{equation}
with $x$ real. Physically, this corresponds to consider the wave function $\psi_n$ in Fourier (Bloch) domain, rather than in physical space. Clearly, for a localized eigenstate $\psi_n$ of $H$, i.e. $\sum_n | \psi_n|^2 < \infty$, the series on the right hand side of Eq.(\ref{S10}) converges and $\psi(x)$ is a regular continuous function and periodic in $x$ with $2 \pi$-period. Conversely, if $\psi_n$ is an extended state of $H$, the series on the right hand side of Eq.(\ref{S10}) is not convergent, and $\psi(x)$ is a $2 \pi$-periodic generalized function, i.e. it should be regarded  as a superposition of $\delta$-Dirac  combs. Physically, in a quasi crystal an extended state can be written as $\psi_n=\sum_l A_l \exp(iq_l n)$, i.e. as a superposition of plane waves of amplitudes $A_l$ and wave numbers $q_l$, with $\sum_l |A_l|^2 < \infty$. This is because a plane wave with the Bloch wave number $q_0$ in the periodic lattice is scattered off by the incommensurate potential $V_n$ to yield a series of plane waves with wave numbers $q_l$ that differ by $q_0$ by integer multiples than $2 \pi \alpha$. Assuming $\psi_n=\sum_l A_l \exp(iq_l n)$,
the series on the right hand side of Eq.(\ref{S10}) can be readily calculated and reads $\psi(x)= 2 \pi \sum_{n,l} A_l \delta(x+q_l-2 \pi n)$, which describes s superposition of Dirac combs.\\
In terms of the function $\psi(x)$, either a regular or a generalized function, Eq.(\ref{S9}) is equivalent to the following difference equation
\begin{equation}
\psi(x) [1+i W(x)]= \exp(2 \epsilon -2 i \theta) [1-i W(x- 2 \pi \alpha)] \psi(x-2 \pi \alpha)  \label{S11}
\end{equation}
where we have set
\begin{equation}
W(x)= \sum_n W_n \exp(inx)=-\frac{E}{V}+ \frac{2}{V} \cos x. \label{S12}
\end{equation}
After introduction of the auxiliary function
\begin{equation}
u(x) \equiv  [1-iW(x)] \psi(x), \label{S13}
\end{equation}
 Eq.(\ref{S11}) is equivalent to the following one
\begin{equation}
u(x- 2 \pi \alpha)=g(x) u(x) \label{S14}
\end{equation}
where we have set
\begin{equation}
g(x) \equiv \exp(-2 \epsilon + 2 i \theta) \frac{1+iW(x)}{1-iW(x)}. \label{S15}
\end{equation}
We note that Eqs.(\ref{S14}) and (\ref{S15}) can be formally interpreted as the Floquet eigenvalue equation obtained from the evolution dynamics, over one period, of a quantum particle with a kinetic energy operator linear in particle momentum and periodically kicked by the potential $\mathcal{V}(x)=2 \;{\rm atan} W(x)$, describe by the following Schr\"odinger equation for the wave function ${\Psi}(x,t)$
\begin{equation}
 i \frac{\partial \Psi}{\partial t}= - 2 \pi  i \alpha \frac{\partial \Psi}{\partial x} + \mathcal{V}(x) \left( \sum_n \delta (t-n) \right) \Psi. \nonumber
\end{equation}
The quasi energy $\mu$ is related to $\theta$ and $\epsilon$ by the relation $\mu=  -2(\theta + i \epsilon)$. Hence, like in the Hermitian case, we have a correspondence between the spectral properties of the NH  quasi crystal and the dynamical properties of a NH periodically-kicked system. As compared to the Maryland model, in its NH extension the kicked potential $\mathcal{V}$ and quasi energy $\mu$ become complex. The advantage of the dynamical formulation of the spectral problem is that Eq.(\ref{S14}) can be solved in an exact form. To this aim, we have to distinguish two cases, depending on whether $\psi_n$ is a localized or an extended wave function of $H(\theta, \epsilon)$, i.e. $u(x)$ a regular or a generalized function of $x$.\\

\subsection{Localized states}
For a localized state, $\sum_n |\psi_n|^2 < \infty$ and thus $u(x)$ is a continuous and $2 \pi$-periodic function of $x$. Since we require $L(E)>0$ for a localized state,  
$E$ does not belong to $\Gamma$, so that $[1-iW(x)]$ is a non-vanishing function on the real $x$ axis. Therefore, from Eq.(\ref{S15}) it follows that also $g(x)$ is a continuous and $2 \pi$-periodic function on the real $£x$ axis. For an irrational $\alpha$, Eq.(\ref{S14}) can be satisfied provided that the following solvability condition is met
\begin{equation}
\int_{0}^{2 \pi} dx \log | g(x)|=0. \label{S16}
\end{equation}
In fact, from Eq.(\ref{S14}) by iteration one has
\begin{equation}
u(x- 2 \pi \alpha n)= \left( \prod_{l=0}^{n-1} g(x-2 \pi \alpha l) \right) u(x). \label{S17}
\end{equation}
Since $ \alpha$ is irrational, one can always find an integer $n$, as large as we wish, such that $2 \pi \alpha n$ is infinitesimally close to an integer multiple than $ 2 \pi$, i.e. $ x-2 \pi \alpha n \simeq x$ mod $ 2 \pi$ with an accuracy as high as we wish for large enough $n$. Due to continuity and periodicity of $u(x)$ this means that the product term on the right hand side of Eq.(\ref{S17}) should converge to one, i.e. 
\begin{equation}
 \prod_{l=0}^{n-1} g(x-2 \pi \alpha l) \simeq 1 \label{S18}
\end{equation}
for $n$ large and $2 \pi \alpha n$ close to an integer multiple than $ 2 \pi$. Taking the log of both sides of Eq.(\ref{S18}), after multiplying both sides by $1/n$ one obtains
\begin{equation}
 \frac{1}{n} \sum_{l=0}^{n-1} \log g(x-2 \pi \alpha l) \simeq 2 \pi i s /n \label{S19}
\end{equation}
with $s$ in an arbitrary integer. From Weyl$^{\prime}$s equidistribution theorem and properties of irrational rotations, for large $n$ the term on the left hand side of Eq.(\ref{S19}) is independent of $x$ and can be approximated by an integral, i.e.
\[  \frac{1}{n} \sum_{l=0}^{n-1} \log g(x-2 \pi \alpha l) \simeq  \frac{1}{2 \pi} \int_0^{2 \pi} dx \log g(x) \]
so that
\begin{equation}
\frac{1}{2 \pi} \int_0^{2 \pi} dx \log g(x)  \simeq 2 \pi i s /n. \label{S20}
\end{equation}
In order to satisfy Eq.(\ref{S20}), the solvability condition Eq.(\ref{S16}) must be met.  Using Eqs.(\ref{S12}) and (\ref{S15}), the solvability condition Eq. (\ref{S16}) can be written as 
\begin{equation}
{\rm Re} (I_+-I_-)= 2 \epsilon,  \label{S21}
\end{equation}
where we have set
\begin{equation}
I_{\pm} \equiv \frac{1}{2 \pi} \int_0^{2 \pi} dx \log \left(E \pm i V -2 \cos x \right). \label{S22}
\end{equation}
The integrals $I_{\pm}$ defined by Eq.(\ref{S22}) can be analytically computed for arbitrary complex energy $E$ by a contour integral method outlined in Appendix B of Ref. \cite{r51}. After introduction of the two complex angles $\varphi_{\pm}$ defined by the relations
\begin{equation}
\cos \varphi_{\pm} \equiv \frac{E \pm i V}{2} \label{S23}
\end{equation}
with the constraint ${\rm Im}(\varphi_{\pm}) < 0$, one has $I_{\pm}=i \varphi_{\pm}$. Hence Eq.(\ref{S21}) reads
\begin{equation}
{\rm Im} (\varphi_--\varphi_+)=2 \epsilon \label{S24}
\end{equation}
with the constraint
\begin{equation}
{\rm Im} (\varphi_+) < {\rm Im} (\varphi_-)<0. \label{S25}
\end{equation}
Equations (\ref{S23}) and (\ref{S24}) define the loci of energies $E$ in complex plane  corresponding to exponentially-localized eigenstates $\psi_n$ of $H(\epsilon, \theta)$. Note that, as expected for an incommensurate potential, the energy spectrum is independent of the phase $\theta$. To provide an analytical form of the energy spectrum, let us note that Eq.(\ref{S24}) is satisfied by letting $\varphi_{-}- \varphi_+=2 i \epsilon +2 \omega$, with $\omega$ arbitrary real number. Moreover, using prosthaphaeresis formulas and Eq.(\ref{S23}), one has
\begin{widetext}
\begin{eqnarray}
\cos \varphi_+ + \cos \varphi_- & = & 2 \cos \left( \frac{\varphi_+ + \varphi_-}{2} \right) \cos \left( \frac{\varphi_+ - \varphi_-}{2} \right)=E  \label{S26}\\
\cos \varphi_+ - \cos \varphi_- & = & -2 \sin \left( \frac{\varphi_+ + \varphi_-}{2} \right) \sin \left( \frac{\varphi_+ - \varphi_-}{2} \right)=iV \label{S27}
\end{eqnarray}
\end{widetext}
and thus
\begin{eqnarray}
\cos \left( \frac{\varphi_+ + \varphi_-}{2} \right) & = &  \frac{E}{2 \cos (\omega + i \epsilon)}  \label{S28}\\
\sin \left( \frac{\varphi_+ + \varphi_-}{2} \right) & = &  \frac{iV}{2 \sin (\omega + i \epsilon)} \label{S29}
\end{eqnarray}
where we used $(\varphi_{-}- \varphi_+)/2= \omega+ i \epsilon$. Finally, taking the square of both sides of Eqs.(\ref{S28},\ref{S29}) and summing the terms so obtained, one has
\begin{equation}
\frac{E^2}{4 \cos^2 (\omega +i \epsilon)}- \frac{V^2}{4 \sin^2 (\omega + i \epsilon)}=1 \nonumber
\end{equation}
i.e.
\begin{equation}
E(\omega)= \sqrt{4 \cos^2 (\omega +i \epsilon)+ V^2 \frac{ \cos^2 (\omega + i \epsilon)} { \sin^2 (\omega + i \epsilon)}} \label{S30}
\end{equation}
 which defines in complex plane the parametric curve $E=E(\omega)$ of point spectrum of $H(\theta, \epsilon)$ [Eq.(5) in the main text]. Note that only the portions of the curve that satisfy Eq.(\ref{S25}) can be considered to belong to the point spectrum of $H(\theta,\epsilon)$. Also, in Eq.(\ref{S30}) we should limit to consider only the square root branch with energy in the upper half complex plane, i.e. with ${\rm Im}(E) \geq 0$. Such a condition readily follows from the fact that, for $ \epsilon>0$, the imaginary part of the potential $V_n=V \tan ( \pi \alpha n + \theta + i \epsilon)$ is non-negative at any lattice site $n$, which implies ${\rm Im}(E) \geq 0$.\\
 \begin{figure*}
\includegraphics[width=18cm]{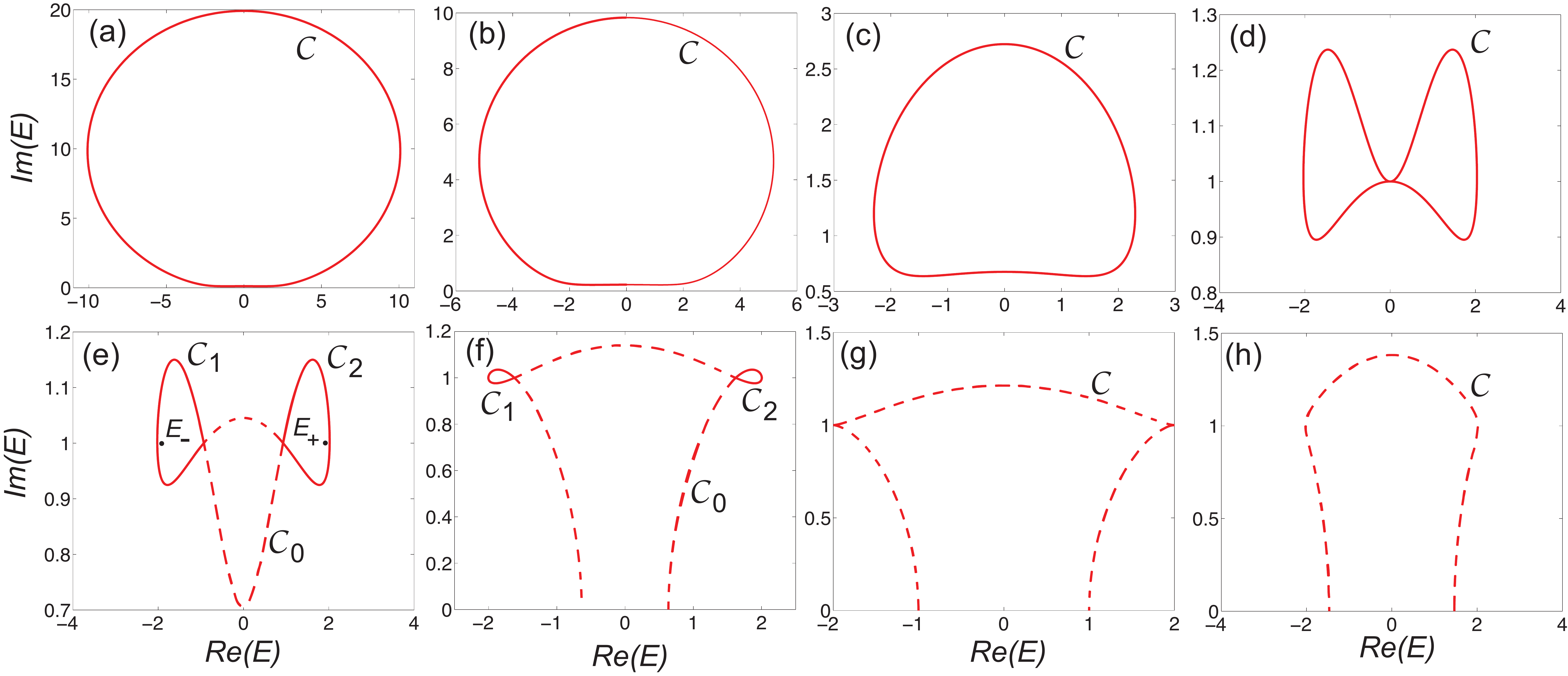}
\caption{(Color online) Behavior of the curve $E=E(\omega)$ in the upper half of complex energy plane, defined by Eq.(\ref{S30}), for $V=1$ and for a few increasing values of $\epsilon$: (a) $\epsilon=0.05$, (b) $\epsilon=0.1$, (c) $\epsilon=0.3$, (d) $\epsilon=\epsilon_1=0.4407$, (e) $\epsilon=0.46$, (f) $\epsilon=0.5$, (g) $\epsilon=\epsilon_2=0.5306$, (h) $\epsilon=0.6$. The solid lines show the branches of the curve $E(\omega)$ where the condition (\ref{S25}) is satisfied, whereas the dashed lines show the branches of the curve $E(\omega)$ where condition (\ref{S25}) is not satisfied. In (e) and (f) the sub-loops $\mathcal{C}_1$ and $\mathcal{C}_2$ encircle the points $E_{\pm}=\pm 2 +iV$.} \label{FigS1}
\end{figure*}
  Typical shapes of the curve $E=E(\omega)$ for increasing values of $\epsilon$ are shown in Fig.\ref{FigS1}. Three cases must be considered.\\
(i) For $\epsilon < \epsilon_1$, with
\begin{equation}
\epsilon_1 \equiv \frac{1}{2} {\rm arsinh}(V)   \label{S31}
\end{equation}
the curve $E(\omega)$ describes a closed loop $\mathcal{C}$ without self-intersections [Figs.\ref{FigS1}(a-c)]. All along the curve the condition (\ref{S25}) is satisfied, so that the entire curve $\mathcal{C}$ defines the point spectrum of $H(\theta, \epsilon)$. Interestingly, in the limit $\epsilon \rightarrow 0^+$ the curve $\mathcal{C}$ is well approximated by the circumference centered at $(0,V/ 2 \epsilon)$ of radius $R= V/(2 \epsilon)$. This shows that the NH Maryland model is a non-perturbative extension of the Hermitian Maryland model.\\
(ii) For $\epsilon_1 < \epsilon < \epsilon_2$, with
\begin{equation}
\epsilon_2  \equiv \frac{1}{2} {\rm arcosh} \sqrt{ \frac{2+V^2}{2} + \sqrt{ \left( \frac{2+V^2}{2} \right)^2 -1}} \label{S32}
\end{equation}
the curve $E(\omega)$ describes a closed loop with two self-intersections, which break the curve $\mathcal{C}$ into three sub loops, a central curve $\mathcal{C}_0$, either closed or open, and two outer closed sub loops $\mathcal{C}_1$ and $\mathcal{C}_2$ that encircle the two point $E_{\pm}= \pm 2+iV$ [Figs.\ref{FigS1}(e,f)].  In this case the condition (\ref{S25}) is not satisfied in the central sub loop $\mathcal{C}_0$, so that the point spectrum of  $H(\theta,\epsilon)$ is described by the two outer loops $\mathcal{C}_1$ and $\mathcal{C}_2$. As $ \epsilon$ approaches $\epsilon_2$ from below, the loops $\mathcal{C}_{1}$ and $\mathcal{C}_2$ shrink toward the two points $E_{\pm}$.\\
(iii) For $\epsilon> \epsilon_2$ the curve $E(\omega)$ describes an open loop $\mathcal{C}$ [Fig.\ref{FigS1}(h)], however the condition (\ref{S25}) is not satisfied nowhere on $\mathcal{C}$. Therefore in this case the point spectrum of $H(\theta, \epsilon)$ is empty.\
\stef{Finally, we note that, for a given eigenenergy $E$ defined by Eq.(\ref{S30}), the corresponding eigenfunction $ \psi(x)$ in Fourier space [Eq.(\ref{S10})] is given by $\psi(x)= u(x)/[1-iW(x)]$, where the $2 \pi$-periodic function $u(x)$ satisfy the difference equation (\ref{S14}), which can found as a Fourier series extending the procedure outlined in \cite{r63b}. To this aim, let us write Eq.(\ref{S27}) in the equivalent form $u(x-2 \pi \alpha)= u(x) \exp \{ -i [i \log g(x)] \}$. A solution to this equation can be found by the Ansatz $u(x)=\exp [-i \Omega_0 x +i P(x) ]$ with $P(x+2 \pi)=P(x)$. The real term $\Omega_0$ and the $2 \pi$-periodic function $P(x)$ are then obtained from the condition $2 \pi \alpha \Omega_0+P(x-2 \pi \alpha)=P(x)-i \log g(x)$, which can be readily solved after Fourier series expansion. The final form for $u(x)$ reads explicitly
\[
u(x)=\exp \left( -i \Omega_0 x- \sum_{n \neq 0} \frac{ \pi \alpha \Omega_n }{\sin (\pi \alpha n)} \exp[in(x+ \pi \alpha)]   \right)
\] 
where we have set
\[
\Omega_n \equiv -\frac{i}{4 \pi^2 \alpha} \int_{-\pi}^{ \pi} dx \left\{ \log g(x) \right\} \exp(-inx).
\]
Note that, since $\int_0^{2 \pi} dx |g(x)|=0$, the coefficient $\Omega_0$ is real.
}


\subsection{Extended states}
According to the Lyapunov exponent analysis presented in Appendix A, the energy $E$ of an extended state $\psi_n$ should belong to the segment $\Gamma$ of the complex energy plane, i.e. there exists a real number $\omega$ in the range $(0, 2 \pi)$ such that 
\begin{equation}
E=iV+2 \cos \omega. \label{S33}
\end{equation}
Note that, since $1-iW(x)$ vanishes at $x=\omega$, the recurrence equation (\ref{S14}) should be understood as 
\begin{equation}
u(x-2 \pi \alpha) [1-iW(x)]=u(x)[1+i W(x)] \exp( -2 \epsilon+2 i \theta). \label{S34}
\end{equation}
Clearly, for irrational $\alpha$ Eq.(\ref{S34}) cannot be satisfied for any non-singular function $u(x)$, unless $u(x)$ identically vanishes for any $x$. In fact, if $u(x)$ were a regular function, from Eq.(\ref{S34}) at $x=\omega$ one obtains $u(\omega)=0$, and thus recursively $u(\omega-2 \pi \alpha l)=0$ for any integer $l \geq 0$. For irrational $\alpha$, as $l$ varies the sequence $\omega-2 \pi \alpha l$ mod $ 2 \pi$ is dense in the interval $(0, 2 \pi)$, and thus one should have $u(x)=0$ almost everywhere in this range. Such a result is consistent with the circumstance discussed above that, when $E$ belongs to the continuous spectrum of $H( \theta, \epsilon)$, $\psi_n$ is an extended state in physical space, the series $\sum_n | \psi_n|^2$ is not convergent and thus the functions $\psi(x)$ and $u(x)$ in Fourier domain, defined by Eq.(\ref{S10}) and Eq.(\ref{S13}), should be regarded as generalized functions. Let us then search for a solution to Eq.(\ref{S34}) as a superposition of $\delta$-Dirac combs with incommensurate periods $ 2 \pi$ and $2 \pi \alpha$, i.e. 
\begin{equation}
u(x)=\sum_{l} U_l \sum_n \delta(x-\omega-2 \pi \alpha l - 2 \pi n) \label{S35}
\end{equation}
with $\sum_l |U_l|^2 < \infty$, which necessarily implies $U_l \rightarrow 0$ as $l \rightarrow \pm \infty$. Substitution of the Ansatz (\ref{S35}) into Eq.(\ref{S34}) yields the following recurrence relation for the amplitudes $U_l$ of Dirac combs
\begin{equation}
U_{l-1} (1-iG_l)= \exp(2i \theta -2 \epsilon) (1+iG_l)U_l \label{S36}
\end{equation}
where we have set
\begin{equation}
G_l=W(x=\omega+2 \pi \alpha l). \label{S37}
\end{equation}
Taking into account that $(1-iG_l)=0$ for $l=0$, the solution to the recurrence equation (\ref{S37}) reads
\begin{equation}
U_l= \left\{
\begin{array}{ll}
0 & l  \geq 0 \\
1 & l=-1 \\
\exp(-2 \epsilon+2 i \theta) \frac{1+iG_{l+1}}{1-iG_{l+1}} U_{l+1} & l<-1
\end{array}
\right. \label{S38}
\end{equation}
i.e.
\begin{equation}
U_l= \left\{
\begin{array}{ll}
0 & l  \geq 0 \\
1 & l=-1 \\
\prod_{n=1}^{-l-1} \exp(-2 \epsilon+2 i \theta) \frac{1+iG_{-n}}{1-iG_{-n}}  & l<-1 .
\end{array}
\right. \label{S39}
\end{equation}
To study the asymptotic behavior of $U_l$ as $l \rightarrow -\infty$, let $I_l=-(1/l)\log |U_{-l}|$ for $l \geq 1$; in the large $l$ limit one has
\begin{widetext}
\begin{equation}
I_l=2 \epsilon - \frac{1}{l} \sum_{n=1}^{l-1} \log \left| \frac{1+iG_{-n}}{1-iG_{-n}} \right| \rightarrow 2 \epsilon- \frac{1}{2 \pi} \int_0^{2 \pi} dx \log \left| \frac{1+i W(x)}{1-iW(x)} \right| \label{S40}
\end{equation}
\end{widetext}
thus the condition $U_l \rightarrow 0$ as $l \rightarrow - \infty$ requires that $I_{\infty}>0$, i.e.
\begin{equation}
\frac{1}{2 \pi} \int_0^{2 \pi} dx \log \left| \frac{V+i V W(x)}{V-i VW(x)} \right| < 2 \epsilon \label{S41}
\end{equation}
Since  $|V+iV W(x)|=| 2 \cos \omega -  2 \cos x +2iV|$ and  $|V-iV W(x)|=| 2 \cos \omega -  2 \cos x|$, taking into account that
\begin{eqnarray}
\frac{1}{2 \pi} \int_0^{2 \pi} dx \log | 2 \cos \omega -2 \cos x |=0  \;\;\;\;  \\
\frac{1}{2 \pi} \int_0^{2 \pi} dx \log | 2 \cos \omega -2 \cos x+2iV  |=- {\rm Im} (\varphi) \nonumber \label{S42}
\end{eqnarray}
with the complex angle $\varphi$ defined by the relation
\begin{equation}
 \cos \varphi = i V + \cos \omega \; \; \; \; \left( {\rm Im}(\varphi)<0 \right) \label{S43}
\end{equation}
the condition $ \sum_l |U_l|^2 < \infty$ is satisfied provided that
\begin{equation}
\left| {\rm Im} (\varphi) \right|< 2 \epsilon. \label{S44}
\end{equation}
Equation (\ref{S44}) determines the interval of energies $E(\omega)=iV+2 \cos \omega$ on the segment $\Gamma$ corresponding to the continuous spectrum of $H(\theta, \epsilon)$. To provide analytical results, let us 
set $\varphi=\varphi_R+i \varphi_I$ ($\varphi_{R}$ and $\varphi_I$ are the real and imaginary parts of the complex angle $\varphi$). From Eq.(\ref{S43}), after elimination of $\varphi_R$ one readily obtains that $\varphi_I$ is given by
\begin{widetext}
\begin{equation}
| \varphi_I|= {\rm arcosh} \sqrt{\frac{1+V^2+\cos^2 \omega}{2}+ \sqrt{\left( \frac{1+V^2+\cos^2 \omega}{2} \right)^2 -\cos^2 \omega} } \label{S45}
\end{equation}
\end{widetext}
From the above equation, it readily follows that, as $\omega$ varies in the range $(0, 2 \pi)$, $|\varphi_{I}|$ is bounded above and below as follows
\begin{equation}
2 \epsilon_1 \leq | \varphi_I| \leq 2 \epsilon_2 \label{S46}
\end{equation}
where $\epsilon_{1,2}$ are defined by Eqs.(\ref{S31}) and (\ref{S32}), and where the boundary $2 \epsilon_2$ ($2 \epsilon_1)$ is attained at $\omega=0$ ($\omega= \pi/2$). 
To determine the continuous spectrum of $H(\theta,\epsilon)$, we thus have to distinguish three cases.\\
 (i) For $\epsilon< \epsilon_1$, Eq.(\ref{S44}) is never satisfied. Hence the continuous spectrum of $H(\theta, \epsilon)$ is empty and there are not extended states for $\epsilon< \epsilon_1$.\\
 (ii) For $\epsilon_1 <\epsilon< \epsilon_2$, Eq.(\ref{S44}) is satisfied for $|\omega|< \omega_0$, where $\omega_0$ is defined by the relation
\begin{equation}
\cos^2 \omega_0 = \cosh^ 2(2 \epsilon) \left\{ 1-\frac{V^2}{\sinh^2(2 \epsilon)} \right\}.
\end{equation}
Hence, for $\epsilon_1 < \epsilon < \epsilon_2$ the continuous spectrum of $H(\theta, \epsilon)$ is given by the interval of energies $E=iV+2 \cos \omega$ with $|\omega|< \omega_0$.\\
(iii) For $\epsilon> \epsilon_2$, Eq.(\ref{S44}) is satisfied for any $\omega$, so all energies on the segment $\Gamma$ of complex plane belong to the continuous spectrum of $H(\theta, \epsilon)$.\\
\stef{Finally, we note that in physical space the wave function $\psi_n$ of the extended state with eigenenergy $E=iV+2 \cos \omega$ is readily obtained from Eq.(\ref{S10}) as $\psi_n=(1 /2 \pi) \int_{-\pi}^{\pi} dx \psi(x) \exp(-inx)$. From Eqs.(\ref{S13}) and (\ref{S35}), after some straightforward calculations one obtains 
\[
\psi_n= \frac{V}{4 \pi i} \sum_l U_l \frac{ \exp[-in(2 \pi \alpha l- \omega)]}{\cos \omega- \cos( \omega- 2 \pi \alpha l)}.
\]
}


\section{Winding numbers, spectral problem in Fourier space and geometry of spectral rotations}
Let us consider  a sequence of rational numbers $\alpha_l=p_l/q_l$, with $q_{l+1} \geq q_l$ and $p_{l+1} \geq p_l$, that is an approximant to the irrational number $\alpha$, i.e. $\alpha=\lim_{l \rightarrow \infty} \alpha_l$. For example, when $\alpha$ is the inverse of the golden mean, i.e. $\alpha=(\sqrt{5}-1)/2$, the sequence $\alpha_l$ is given in terms of Fibonacci numbers $p_l=0,1,1,2,3,5,8,13,21,34,55, 89, 144, ...$ with $q_l=p_{l+1}$. For large enough $l$, let us consider a lattice comprising $L=q_l$ sites under periodic (ring) boundary conditions $\psi_{n+L}=\psi_n$, and let $H(\theta /L,\epsilon)$ be the $L \times L$ matrix Hamiltonian corresponding to the potential
\begin{equation}
V_n=V \tan ( \pi \alpha_l n + \theta /L +i \epsilon). \label{S48}
\end{equation}
Note that $V_n$ is periodic with respect to $n$ with period $L$, i.e. $V_{n+L}=V_n$. For a given base energy $E_B$ that does not belong to the spectrum of $H(\theta /L,\epsilon)$, one can introduce the following winding number
\begin{equation}
w(E_B)= \lim_{L \rightarrow \infty} \frac{1}{2 \pi i} \int_0^{\pi} d \theta \frac{\partial }{\partial \theta} \log \left\{ \det \left( H \left(  \frac{\theta}{L}, \epsilon \right)-E_B \right) \right\}. \label{S49}
\end{equation}  
The winding number $w(E_B)$ has a very simple geometrical interpretation in terms of spectral rotations of the eigenvalues of $H ( \theta /L, \epsilon)$. In fact, let us indicate by $E_1( \theta )$, $E_2( \theta )$, ...,  $E_L( \theta)$ the $L$ eigenvalues of $H ( \theta /L, \epsilon)$. Note that each energy $E_l(\theta)$ is a continuous function of $\theta$; owing to the periodicity of the potential $V_n$, as $\theta$ spans from $\theta=0$ to $\theta=\pi$ the spectrum should reproduce itself at the end of the cycle. Indicating by $\Delta \varphi_l$ the angle spanned by the vector $E_l(\theta)-E_B$  as $\theta$ varied from $0$ to $\pi$ [see Fig.5(c)], one has
\begin{equation}
w_B=\frac{1}{2 \pi} \sum_{l=1}^{L} \Delta \varphi_l \label{S50}
\end{equation}
The above equation indicates that the winding number is related to the flow (spectral rotations) of the eigenenergies of $H(\theta/L,\epsilon)$ in complex plane as $\theta$ continuously varies from $0$ to $\pi$. To capture the geometrical properties of spectral rotations, it is worth considering the spectral problem in Fourier (Bloch) space, rather than in physical space, by introducing the unitary transformation 
\begin{equation}
\phi_n=  \frac{1}{\sqrt{L}} \sum_{l=1}^{L}  \psi_l \exp(2  i \pi \alpha n l +2 i \theta n/L) \label{S51}
\end{equation}
with the inverse relation
\begin{equation}
\psi_n= \frac{1}{ \sqrt{L}} \sum_{l=1}^{N} \phi_l \exp(-2 \pi i \alpha l n- 2 i \theta n/L) \label{S52}
\end{equation}

 \begin{figure*}
\includegraphics[width=18cm]{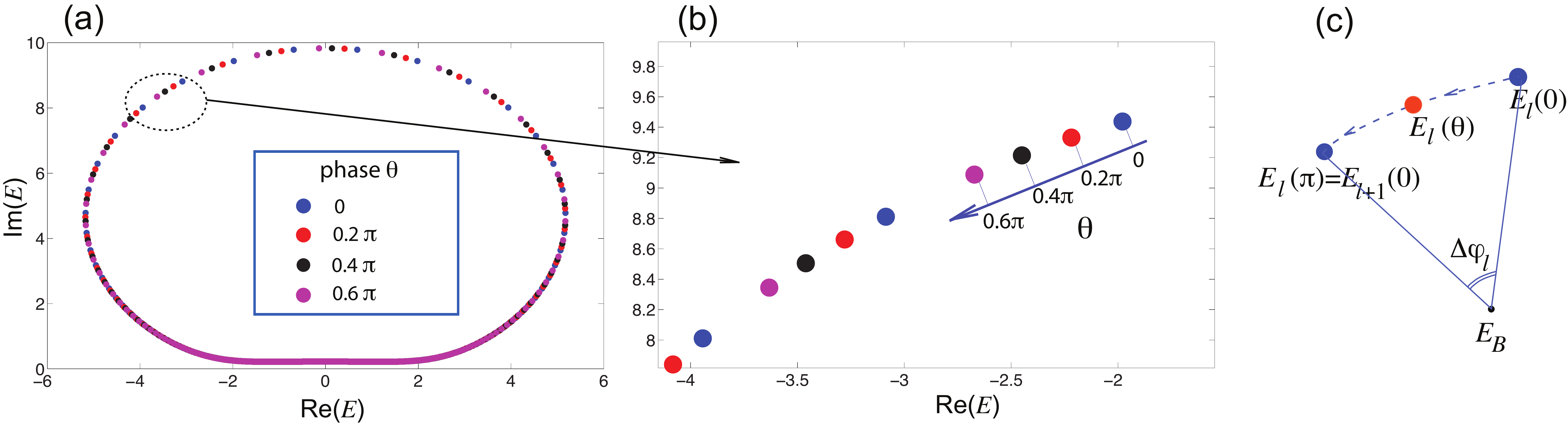}
\caption{ \stef{(Color online) (a) Energy spectrum $E_l(\theta)$ in complex plane of $H(\theta /L, \epsilon)$ under periodic boundary conditions for a few increasing values of the angle $\theta$ and for $\epsilon=0.2$, $V=1$ and $L=233$ (corresponding to the rational approximation $\alpha \simeq 144/233$ of the inverse of the golden mean). Panel (b) shows an enlargement of (a) to illustrate the flow of the eigenvalues $E_l(\theta)$ as the angle $\theta$ increases. As $\theta$ varies continuously from $\theta=0$ to $\theta= \pi$, the $l$-th eigenvalue $E_l(\theta)$ of $H(\theta /L, \epsilon)$ flows along a short path on the loop to coincide with the next eigenvalue $E_{l+1}(\theta=0)$. Correspondingly, the vector $E_l(\theta)-E_B$ in complex plane spans an angle $\Delta \varphi_l$ with respect to the base energy $E_B$.}}
\end{figure*}
Since $\psi_n$ satisfies the  ring boundary conditions $\psi_{n+L}=\psi_n$, from Eq.(\ref{S51}) it readily follows that $\phi_n$ satisfies the ring boundary conditions with an applied magnetic flux $2 \theta$, i.e.
\begin{equation}
\phi_{n+L}= \phi_n \exp(2 i \theta). \label{S53}
\end{equation}
The spectral problem in Fourier space representation $\phi_n$ reads
\begin{equation}
E \phi_n= [2 \cos (2 \pi \alpha n)+iV] \phi_n+ V \sum_{l \neq 0}  S_{l} \phi_{n-l} \equiv \tilde{H} \phi_n \label{S54}
\end{equation}
where 
\begin{equation}
S_{l} \equiv \frac{1}{2 \pi} \int_0^{2 \pi} dx \; {\rm tan} \left( \frac{x}{2}+i \epsilon \right) \label{S55}
\end{equation}
are the Fourier coefficients of the complex tan potential. Such coefficients can be readily computed by the Cauchy's residue theorem and read explicitly
\begin{equation}
S_l= \left\{
\begin{array}{ll}
2i[-\exp(-2 \epsilon)]^{|l|} & l \leq -1 \\
i  & l=0 \\
0 & l \geq 1
\end{array}
\right. \label{S56}
\end{equation}
The eigenenergies $E_l(\theta)$ of the Hamiltonian $H( \theta/L, \epsilon)$ can be thus computed from the spectral problem in Fourier space, defined by Eq.(\ref{S54}) with periodic boundary conditions on a ring threaded by a magnetic flux $2 \theta$ [Eq.(\ref{S53})]. Note that the matrix Hamiltonian $\tilde{H}$ in Fourier space, defined by Eq.(\ref{S54}), does not dependent on $\theta$, and describes the hopping dynamics on a lattice with  long-range hopping amplitudes $V S_l$ in an incommensurate sinusoidal potential $\mathcal{V}_n \equiv [2 \cos (2 \pi \alpha n)+iV]$. Interestingly, the hopping $VS_l$ is {\it unidirectional}, i.e. it vanishes for any $l>0$.  While $\tilde{H}$ is independent of $\theta$, the eigenvalues $E_l$ depend on $\theta$ via the boundary condition Eq.(\ref{S53}).\\
 If the eigenvalue $E_l$ corresponds to an extended wave function $\psi_n$ in physical space, then in Fourier space $\phi_n$ is a localized eigenstate, and thus insensitive to the magnetic flux $2 \theta$ threading the ring. This means that any eigenvalue $E_l(\theta)$, corresponding to an extended wave function $\psi_n$ in physical space, is independent of the phase $\theta$ and thus the phase contributions $\Delta \varphi_l$ to the winding number $w(E_B)$, entering in Eq.(\ref{S50}), vanishes. This proves that in the fully delocalized phase of the Maryland model one has $w(E_B)=0$ for any base energy. Interestingly, for an extended eigenfunction $\psi_n$ in physical space (and thus for a localized wave function $\phi_n$ in Fourier space), the unidirectionality of hopping in $\tilde{H}$ ensures that the eigenvalue $E_l$ is a diagonal element of $\hat{H}$, i.e. there is an integer $n=n_l$ such that $E_l=\mathcal{V}_{n_l}=[2 \cos(2 \pi \alpha n_l)+iV]$, in agreement with the analysis of Appendix B [Eq.(\ref{S33}) with $\omega=2 \pi \alpha n_l$]. Moreover, the localized eigenfunction $\phi_n$ in Fourier space is {\it unilateral}, i.e. $\phi_ n=0$ for $n>n_l$, while the explicit form of $\phi_n$ for $n <n_l$ can be readily computed from Eq.(\ref{S54}) by iteration assuming $\phi_{n_l}=1$, i.e.
 \begin{equation}
 \phi_n= \left \{
 \begin{array}{ll}
 \frac{V}{2} \frac{1}{\cos(2 \pi \alpha n_l)-\cos(2 \pi \alpha n)} \sum_{l=1,2,3,...} S_{-l} \phi_{n+l}  & n<n_l \\
 1 & n=n_l \\
 0 & n>n_l
 \end{array}
 \right.
 \end{equation}
Let us now assume that the eigenvalue $E_l(\theta)$ corresponds to a localized wave function $\psi_n$ in physical space, and thus to an extended wave function $\phi_n$ in Fourier space. Such extended wave function is now sensitive to the magnetic flux $2 \theta$ threading the ring, and thus $E_l(\theta)$ should now depend on $\theta$. The eigenvalues $E_l(\theta)$ for localized states $\psi_n$ lie on the closed solid loops ($\mathcal{C}$ or $\mathcal{C}_{1,2}$) of Fig.4. Since such loops are independent of $\theta$, this means that, as $\theta$ is varied, the eigenvalues $E_l(\theta)$ of localized states $\psi_n$ undergo geometric rotations on the loops. Such spectral rotations of eigenvalues induced by a magnetic flux is fully analogous to the scenario found in other non-Hermitian models like in the Hatano-Nelson-Anderson model (see for example Appendix B.1 of Ref.\cite{r1}), and is illustrated in Fig.5. The figure depicts the eigenvalues $E_l(\theta)$ versus $\theta$ of $H( \theta/L, \epsilon)$ in a ring lattice comprising $L=233$ sites, corresponding to the rational approximation $ \alpha \simeq 144/233 $ of the inverse of the golden mean, for $\epsilon=0.2$ (all wave functions $\psi_n$ are exponentially localized). As the magnetic flux $\theta$ is continuously varied from $0$ to $\pi$, each eigenvalue $E_l(\theta)$ moves along the closed loop $\mathcal{C}$ [Fig.5(a) and (b)] and rotates by an angle $\Delta \varphi_l$ with respect to the base energy $E_B$  [Fig.5(c)]. After a full scan of the magnetic flux, from $\theta=0$ to $\theta= \pi$, the overall spectrum is unchanged and each eigenenergy on the loop flows to another eigenenergy, i.e. $E_{l}(\theta= \pi)=E_{l+1}(\theta=0)$, as illustrated in Fig.5(c). Hence, if the base energy $E_B$ is internal to the loop, the spectral rotations of eigenvalues contribute to  the winding number  as $\sum_{l} \Delta \varphi_l=2 \pi$, and thus $w(E_B)=1$.\\


\begin{thebibliography}{31}
\bibitem{nota}
We note that, since the  NH Maryland model does not show the NH skin effect, energy spectra under open or periodic boundary conditions are basically the same in the thermodynamic limit $L \rightarrow \infty$.
\end{thebibliography}
\end{document}